\documentclass{ifacconf}
\usepackage{numbersUncertaintyIFAC}
\usepackage{enumitem}       

\begin{document}
\begin{frontmatter}

\pdfminorversion=4              

\title{ On the validity of using the delta method for calculating the uncertainty of the predictions from an overparameterized model \thanksref{footnoteinfo}}

\thanks[footnoteinfo]{	This work is supported by Sweden's innovation agency, Vinnova, through project iQDeep (project number 2018-02700).}

	\author[Linkan]{Magnus Malmstr{\"o}m},    
\author[UU]{Isaac Skog},               
\author[Linkan]{Daniel Axehill},   
\author[Linkan]{Fredrik Gustafsson}  

\address[Linkan]{Link{\"o}ping University, Sweden , (firstname.lastname@liu.se)}
\address[UU]{Uppsala University, Sweden, (firstname.lastname@angstrom.uu.se)}  

\begin{abstract}                
The uncertainty in the prediction calculated using the delta method for an overparameterized (parametric) black-box model is shown to be larger or equal to the uncertainty in the prediction of a  canonical (minimal) model. Equality holds if the additional parameters of the overparameterized model do not add flexibility to the model. 
As a conclusion, for an overparameterized black-box model, the calculated uncertainty in the prediction by the delta method is not underestimated.
The results are shown analytically and are validated in a simulation experiment where the relationship between the normalized traction force and the wheel slip of a car is modelled using e.g., a neural network.
\end{abstract}


\end{frontmatter}

\section{Introduction}
\noindent
This paper investigates how overparameterization affects
the, via the delta method, calculated uncertainty in the prediction from a parametric black-box model, such as a neural network (\nn).
To be able to use a model in a safety-critical application, such as medical image recognition or autonomous driving, it is important to be able to quantify the uncertainty in the predictions of the model, \citep{paleyes2020challenges}.  
For \nn{s}, there are numerous methods to quantify the uncertainty in the predictions, \citep{gawlikowski2021survey}.
Here, the delta method, \citep{Hannelore2011,malmstrom2021} is an example of such a method.
	It relies on a two-step procedure. Firstly, to compute the uncertainty of the parameters in the black-box model, and secondly to propagate, through linearization, the uncertainty in the parameters to uncertainty in the output.  Hence, it is a method based on identifying a distribution for the parameters of the black-box model.
 The method shares similarities to the Laplacian approximation of Bayesian \nn{s}, \citep{immer2021improving}.
Broadly speaking, methods to quantify the uncertainty in the predictions can be separated into two categories. The first category of methods is based on creating ensembles of predictions, \citep{lakshminarayanan2016simple, Gal2015a,malmstrom2021}. The second category of methods is based on learning the uncertainty by modifying the \nn structure or the cost function, \citep{Kendall2017, Blundell2015,izmailov2021dangers}.

For black-box models, such as \nn{s}, a common design choice is to use an overparameterized model to guarantee that the model is flexible enough to describe the true system. 
In the literature, the problem of quantifying the uncertainty for overparameterized models has previously been studied, see  \citep{hjalmarsson1995composite, Stoica2001, Pintelon1996,stoica1989reparametrization}.
For example,  independently of how a system has been overparameterized, \citep{Pintelon1996} shows that the uncertainty in the learned (estimated) parameters is the same. 
In this paper, the contribution is to show, that using the delta method to compute the uncertainty in the prediction of a black-box model, the uncertainty is always larger for an overparameterized model compared to the calculated uncertainty from a model with minimum flexibility required to describe the true system. However,  the calculated uncertainty in the prediction is the same if the added parameters of the overparameterized model do not add flexibility to the model.  

\section{Problem formulation}
\noindent This paper will consider regression problem with least-squares loss function, and investigate how the use of overparameterization affects the prediction uncertainty calculated by the delta method.

\subsection{Signal model and likelihood}
\noindent Consider a mapping between an input $\xn \in \mathbb{R}^{n_x}$ to some output $f^\ast(\xn)\in \! \mathbb{R}$. Here $n_x$ is the dimension of the input.
A parametric black-box model 
$f(\xn;\bmtheta)$
is used to model this relationship, where $\bmtheta \in \mathbb{R}^{n_{\bmtheta}}$ is the model parameters. 
Assume a scalar measurement $y_n \in \! \mathbb{R}$ of the output $f^\ast(\xn)$ given by 
\begin{align}
y_n = f^\ast(\xn) + e_n,
\end{align}
where $e_n$ is i.i.d. measurement noise.
The parameters of the black-box model are learned from the measurements $y_n$ by minimizing a loss function $V_N(\bmtheta)$, i.e.,  
\begin{align}
\bmthetahat &= \argmin_{\bmtheta} V_N(\bmtheta). \label{eq:paramest}
\end{align}  
The least-squares loss function
\begin{align} \label{eq:lossfunction}
V_N(\bmtheta) = \sum_{n=1}^{N} || y_n - f(\xn;\bmtheta)||^2
\end{align}
is a common choice for regression problems. Here $N$ denotes the number of data points. If the noise $e_n$ has a Gaussian distribution, and loss function \eqref{eq:lossfunction} is used, the estimate in \eqref{eq:paramest} is the maximum likelihood estimate (\mle).

\subsection{Neural network model structure}
\noindent 
A fully connected \nn with $L$ layers can be written as
\begin{subequations}  \label{eq:deepnn}
	\begin{align}
	&\h^{(0)} = \x, \label{eq:deepnn1} \\
	& \aaa^{(l+1)} =   \begin{bmatrix} \h^{(l)} & 1
	\end{bmatrix}^\top \W^{(l)} , \quad l = 0, \ldots , L-1,\\
	&\h^{(l)}=  \sigma \big(\aaa^{(l)} \big), \quad l = 1, \ldots , L-1, \label{eq:deepnn2} 
	\end{align}
	where  $\sigma()$ denotes the activation function. The latent variable $\aaa^{(l)}$ containing the value of all the nodes in the $l$'th layer of the \nn, and $\h^{(l)}$ denotes the transformation using the activation function of the values in all the nodes in the $l$'th layer of the \nn.   
	Collecting all the weights and biases included in the matrices $W^{(L)}, \ldots ,W^{(0)}$ into the parameter vector using the $\text{Vec}(\cdot)$ function, i.e.,
	\begin{align}
	\bmtheta &\triangleq \begin{bmatrix} \text{Vec}(\W^{(L)})^\top & \hdots & \text{Vec}(\W^{(0)})^\top \end{bmatrix}^\top,
	\end{align} 
	the \nn can be written as a parametric model  
	\begin{align} \label{eq:latent}
	f(\x; \bmtheta) &= \aaa^{(L)}.
	\end{align}
\end{subequations}
\subsection{True system}
\noindent Define a model set $\mathcal{M}^\ast$ as a collection of candidate models, \citep{lennart1999system}. If the model set generated by the black-box model $f(\xn;\bmtheta)$ includes the true system $f^\ast(\xn)$, then there exists a $\bmtheta_0$ such that 
\begin{align}
f(\xn;\bmtheta_0) =f^\ast(\xn).
\end{align}  
In practice, it is difficult to guarantee that $\mathcal{M}^\ast$ includes the true system. However, if a very flexible model, such as a deep \nn is used, it can be assumed that such a $\bmtheta_0$ exists, \citep{liang2016deep}.   
	Hence, for these applications, it is sensible to assume that the model set includes the true system.
For \nn, there are some symmetries in the parametrization that makes $\bmtheta_0$ non-unique.  
These symmetries can be handled similarly to overparameterization, but a description of them is outside of the scope of this paper. Interested readers are referred to \citep{Gene1997}.

If the true system is contained in the model set $\mathcal{M}^\ast$, an estimate of $\bmtheta$ found by solving \eqref{eq:paramest} is, asymptotically in the number of data points, the \mle. This is true whether the noise is Gaussian distributed or not,  \citep{lennart1999system}.   
From here on, throughout this paper, it will be assumed that all model sets include the true system, regardless of if they are generated by an overparameterized model or not.

\subsection{Parameter covariance}
\noindent Under the assumption that the signal-to-noise ratio (\snr) tends to infinity, the \mle gives that the estimated parameters convergence in distribution to
\begin{align} \label{eq:gaussianTheta}
\bmthetahat \sim \mathcal{N}(\bmtheta_0,P^{\bmtheta}_N).
\end{align}
Here $P^{\bmtheta}_N$ is the Cram\'{e}r-Rao lower bound (\crlb). 
For the regression problem in \eqref{eq:lossfunction}, the \crlb is given by
\begin{subequations} \label{eq:crlb_cov}
	\begin{align}
	P_N^{\bmtheta} &= \lambda_N \big(\mathcal{I}^{\bmtheta}\big)^{-1}, \\
	\mathcal{I}^{\bmtheta} & = \sum_{n=1}^{N} \mathcal{I}^{\bmtheta}_n, \\
	\mathcal{I}^{\bmtheta}_n & = E \big[  \psi (\xn;\bmtheta) \psi^\top(\xn;\bmtheta) \big],
	\end{align}
\end{subequations}
where $\mathcal{I}^{\bmtheta}$ denotes the information matrix, the prediction error and the prediction error variance are given by
\begin{subequations}
	\begin{align}
	\epsilon(\xn;\bmtheta) &= y_n- f(\xn;\bmtheta), \\
	\lambda_N &= \frac{1}{N} \sum_{n=1}^{N} \epsilon^2(\xn;\bmtheta),\label{eq:prediction_error_varaince}
	\end{align}
\end{subequations}
and the derivative of the prediction error with respect to the parameters $\bmtheta$
\begin{align}
\psi(\xn;\bmtheta) = - \frac{\partial}{\partial \bmtheta}\epsilon(\xn;\bmtheta) = \frac{\partial}{\partial \bmtheta}f(\xn;\bmtheta).
\end{align}

\subsection{Error propagation using the delta method}
\noindent  
The delta method \citep{Hannelore2011}, is a method to propagate the uncertainty in the parameter to uncertainty in the prediction using a linearization of the model. That is, the covariance of the prediction is given by 
\begin{align} \label{eq:linearize}
P^{f}_N(\x) \triangleq \text{var}\big(f(\x;\bmthetahat)\big)= \psi^\top(\x;\bmtheta) P^{\bmtheta}_N  \psi(\x;\bmtheta) , 
\end{align}
for any input $\x$. 
	It is based on the observation that, if the parameters are Gaussian distributed, close to the parameters $\bmthetahat$, the model can accurately be represented by a linearized local approximation around $f(\x;\bmthetahat)$, \cite{Hannelore2011}.
In system identification, this is a standard method to project the uncertainty of the parameters onto the prediction, see  \citep{Gene1997,rivals2000construction, Papadopoulos2001, Chryssoloiuris1996}. 
Then, the uncertainty in the prediction is given by the norm of the projection, \citep{hjalmarsson2010geometric}.

This paper will study how the prediction uncertainty given by \eqref{eq:linearize} is affected by using an overparameterized model.
To do so, overparameterized models will be separated into two categories:
\begin{enumerate}[label=(\roman*)]
	\item Models where the redundant parameters do not add any flexibility, e.g., when some elements in $\aaa^{(l)}$ can be written as a linear combination of the other elements of $\aaa^{(l)}$. These will be referred to as redundant parameters of Category 1. 
	\item Models where the redundant parameters add flexibility, e.g., when more hidden nodes than necessary are used for an \nn. This will be referred to as redundant parameters of Category 2.   
\end{enumerate}

\section{Overparametized models}
\noindent  
The delta method is based on a linear approximation of a nonlinear model. Hence, to analyze how \eqref{eq:linearize} is affected by overparameterization, consider the linearization 
\begin{align} \label{eq:linear_model}
f(\x; \bmtheta) = \bmvarphi^\top(\x) \bmtheta +d,
\end{align}
where $\bmvarphi(\x)=\psi(\x;\bmtheta)$, and $d$ is the difference between the value of the function and the linear approximation at the linearization point. 
However, $d$ will not influence the uncertainty in the prediction calculated by the delta method, hence assume $d=0$.
Note that for a  model linear in the parameters, the information matrix can be written as    
\begin{subequations} \label{eq:linear_covariance}
	\begin{align}
	\mathcal{I}^{\bmtheta} & = \bmPhi \bmPhi^\top,
	\end{align}
	where
	\begin{align}
	\bmPhi &= \begin{bmatrix}\bmvarphi(\x_1) & \cdots  & \bmvarphi(\x_N) \end{bmatrix}.
	\end{align}
\end{subequations}

A model of minimum flexibility but still flexible enough to describe the true system can be referred to as a canonical model. To make a distinction between whether the model is canonical or overparameterized,  a subindex $c$ is added here.

\subsection{Models with redundant parameters of Category 1}
\noindent As an example that the true input-output relationship is given by 
\begin{align} \label{eq:linear_example}
f^\ast(x) = 1+ 6x+x^2.
\end{align} 
Then one canonical model is given by the regressor $\bmvarphi_{c}^\top(x) = [1,x,x^2]$, while $\bmvarphi^\top(x) = [1,x,x+1,x^2]$ is an overparameterized model with redundant parameters of Category 1.  
Hence, there exists some transformation $T(\bmtheta) = \bmtheta_c$ 
which transforms the parameters of the overparameterized to parameters of the canonical representation model. In the linear case,
\begin{align} \label{eq:transformation_lin}
\bmtheta_c & = T\bmtheta.
\end{align}
Here $T \in \mathbb{R}^{n_{\bmtheta_c}\times n_{\bmtheta}}$ is a transformation matrix where $n_{\bmtheta_c}$ is the number of parameters in the canonical model which is smaller than or equal to $n_{\bmtheta}$.
The case where $n_{\bmtheta_c}=n_{\bmtheta}$ represents when there exist some symmetries in the model, e.g., changing the ordering of the nodes in \eqref{eq:deepnn}. 

For \eqref{eq:crlb_cov} to hold, it is assumed that the information matrix $\mathcal{I}^{\bmtheta}$ is invertible. That is, the data is informative enough, and the model is not overparameterized. 
Assume that the data is informative enough with respect to the model set generated by the canonical model.
For an overparameterized model with redundant parameters of Category 1, where the information matrix is singular, the inverse is replaced by a Moore-Penrose pseudo-inverse denoted with the superscript $+$,  \citep{Stoica2001}, i.e., 
\begin{align} \label{eq:parma_with_psudo}
P_N^{\bmtheta} &= \lambda_N \big(\mathcal{I}^{\bmtheta}\big)^{+},
\end{align}
which also works when there exists a null space in the parameter space.

\begin{thm}\label{th:redundentNonlin}

	Consider a canonical model of the true system $ f^\ast(\x)=\bmvarphi^\top_{c}(\x)  \bmtheta_{c,0} $. Form a new model with more parameters $\bmvarphi^\top(\x)  \bmtheta$ using the (wide) transformation matrix $T$ with full row rank. Then, the models estimated by  \eqref{eq:paramest} will give identical uncertainty in the prediction independently of the choice of $T$, i.e.,
	\begin{align}
	\bmvarphi^\top_{c}(\x_m) P^{\bmtheta_c}_{N} \bmvarphi_{c}(\x_m)=\bmvarphi^\top(\x_m) P^{\bmtheta}_N \bmvarphi(\x_m).
	\end{align}  
\end{thm}

\begin{pf}
	With the transformation given in \eqref{eq:transformation_lin}, the regressors for the overparameterized model can be written as $\bmvarphi^\top(\x_m) = \bmvarphi_{c}^\top(\x_m) T$. Hence, the information matrix for the overparameterized model can be written as  
	\begin{align}
	\mathcal{I}^{\bmtheta} = T^\top \mathcal{I}^{\bmtheta_c} T. 
	\end{align}
	Recall that, if $A$ has full column rank, $B$ has full row rank and $C$ is invertible the pseudo-inverse of their product is 
	\begin{align}
	(ACB)^+ = B^+ C^{-1} A^+. 
	\end{align}
	Let $I_r$ denote the identity matrix of size $r$.
	Using that $ T T^+=I_{n_{\bmtheta_c}}$, the uncertainty in the prediction for the canonical model can be written as   
	\begin{subequations}
		\begin{align}
		\bmvarphi_{c}^\top  (\!\x_m \!) & P^{\bmtheta_c}_{N} \bmvarphi_{c}(\!\x_m \!) 	 \!   =  \\
		&= \lambda_N \bmvarphi_{c}^\top (\!\x_m \!) T T^+\big(\mathcal{I}^{\bmtheta}_c \big)^{-1}  \big(T^\top \big)^+\! T^\top\! \bmvarphi_{c}(\!\x_m \!) \\
		& = \! \lambda_N \! \bmvarphi^\top(\!\x_m \!) \!\big(T^\top \! \mathcal{I}^{\bmtheta_c}  T\big)^{\! +} \! \! \bmvarphi(\!\x_m \!)  \\
		&= \bmvarphi^\top (\!\x_m \!)  P^{\bmtheta}_N \bmvarphi(\!\x_m \!) 
		\end{align}
	\end{subequations}
	This is identical to the uncertainty from \eqref{eq:linearize} which was calculated using an overparameterized model with redundant parameters of Category 1. 
	\begin{flushright}
		\qed
	\end{flushright}
\end{pf}

Similar results have been presented in \citep{Stoica2001}. There it is shown that even though the information matrix for the parameters is singular, a projection of the matrix can give an invertible information matrix. In \citep{Pintelon1996} it is shown that the \crlb is independent of the given overparameterization. 
The main difference in Theorem~\ref{th:redundentNonlin}, is that it is the uncertainty in the prediction that is considered where equivalence is shown to a canonical model.  Thereby, it is shown that the effect of the null space in $\mathcal{I}^{\bmtheta}$ can be neglected. This is a result of the structure from having the same transformation when computing the parameter covariance and propagating the uncertainty.

\subsection{Models with redundant parameters of Category 2}
\noindent Once again, consider the problem to identify a model for the system in \eqref{eq:linear_example}, but now, the overparameterized model has parameters that add unnecessary flexibility, i.e., redundant parameters of Category 2. For example if the regressor  $\bmvarphi^{\top}(x) = [1,x,x^2,x^3]$ is used.  
If the model has redundant parameters of Category 2, the information matrix for the overparameterized model is likely to be invertible.

\begin{thm}\label{th:largerModel} 
	Consider the linear model \eqref{eq:linear_model} and assume that the true system can be described by a canonical model with less flexibility. If both models are estimated using \eqref{eq:paramest} where $n_{\bmtheta}$ is fixed and $n_{\bmtheta} \ll N$, then as the \snr goes to infinity, the uncertainty in the prediction \eqref{eq:linearize} for a canonical model is smaller compared to that of an  overparameterized model, \ie,
	\begin{align} \label{eq:sigVarOverLinpar}
	\bmvarphi_{c}^\top(\x_m) P^{\bmtheta_c}_N \bmvarphi_{c}(\x_m)<\bmvarphi^\top(\x_m) P^{\bmtheta}_N \bmvarphi(\x_m).
	\end{align}
\end{thm}	
\begin{pf}
	Without loss of generality, the overparameterized model can be written in terms of the canonical model such as 
	\begin{subequations}
		\begin{align} \label{eq:OverFuncofCan}
		\bmvarphi^\top(\x_m) &= \begin{bmatrix}  \bmvarphi_{c}^\top(\x_m) & \bmvarphi_{o}^\top(\x_m) \end{bmatrix}, \\
		\bmPhi^\top &= \begin{bmatrix}  \bmPhic^\top & \bmPhio^\top  \end{bmatrix}, \\
		\bmPhi \bmPhi^\top & = \begin{bmatrix}   \bmPhic \bmPhic^\top & \bmPhic \bmPhio^\top \\ \bmPhio \bmPhic^\top & \bmPhio \bmPhio^\top \end{bmatrix}, \label{eq:Larger_model_info}
		\end{align}
	\end{subequations}
	where $\bmvarphi_{o}^\top(\x_m)$ and $\bmPhio^\top$ correspond to the parameters added in the  overparameterized model. Using the block-wise inverse the parameter inverse of the information matrix for the parameters of the overparameterized model is
	\begin{subequations}
		\begin{align} \label{eq:blkInvOver}
		\big( \!\bmPhi \bmPhi^\top \! \big)^{\!-1} \! \! \!= \! \! \begin{bmatrix} \big(\! \bmPhic \bmPhic^\top \! \big)^{\! \!-1} \!  \! \! + \! K^\top R_o K  & -K^\top R_o \\ - R_o K & R_o \end{bmatrix} \! \! ,
		\end{align}
		where 
		\begin{align} \label{eq:simplifyNotation}
			R_o & \!\!= \! \! \big( \! \bmPhio (I_N- R_c) \bmPhio^\top \! \big)^{\!-1} \! \!,\\
			R_c &= \bmPhic^\top  \big(\bmPhic \bmPhic^\top\big)^{-1} \bmPhic, \\
		K & = \bmPhio \bmPhic^\top  \big(\bmPhic \bmPhic^\top\big)^{-1}.
		\end{align}
	\end{subequations}
	Since $n_{\bmtheta} \ll N$ and both the model set of the canonical and overparameterized model include the true system, the prediction error variance is asymptotically the same and equal to the variance of the noise $\lambda_0$, \citep{lennart1999system}.
	Then the variance of the overparameterized model can be written as 
	\begin{align} \label{eq:OverSmallerProof} 
	\lambda_0  & \bmvarphi^\top\! (\!\x_m \!) \big(\bmPhi \bmPhi^\top \big)^{\!-\!1\!} \!\bmvarphi(\!\x_m \!) \! = \! \lambda_0 \bmvarphi_{c}^\top \! (\!\x_m \!) \big(\!\bmPhic \bmPhic^\top \! \big)^{\!-\! 1\!} \! \bmvarphi_{c} (\!\x_m \!) \nonumber \\
&    + \lambda_0 \bmvarphi_{c}^\top(\!\x_m \!) K^\top R_o K \bmvarphi_{c}(\!\x_m \!) \! -\! \lambda_0 \bmvarphi_{c}^\top(\!\x_m \!) K^\top R_o \bmvarphi_{o}(\!\x_m \!)  \nonumber\\
& - \lambda_0 \bmvarphi_{o}^\top (\!\x_m \!)  R_o K \bmvarphi_{c}(\!\x_m \!) + \lambda_0 \bmvarphi_{o}^\top(\!\x_m \!)  R_o \bmvarphi_{o}(\!\x_m \!). 
	\end{align}
	Here \eqref{eq:OverSmallerProof} can be seen as a quadratic optimization problem where the regressor that adds flexibility to the model  $\bmvarphi_{o}^\top(\x_m)$ and $\bmPhio$ are free to choose. Write 
	\begin{subequations}
		\begin{align} 
		\bm{\chi}\top = [ \bmvarphi_{c}^\top(\x_m) K^\top, \bmvarphi_{o}^\top(\x_m) ],
		\end{align} 
		then the optimization problem becomes
		\begin{align} \label{eq:OverSmallerProof_mini}
		\min_{\bm{\chi}} \lambda_0 \bmvarphi_{c}^\top(\!\x_m \!)  \!\big(\bmPhic \bmPhic^\top \big)^{\!-1}\!  \bmvarphi_{c}(\!\x_m \!)  \!+\! \lambda_0 \bm{\chi}^\top \! Q \bm{\chi},  
		\end{align}
		where 
		\begin{align} \label{eq:OverSmallerProof_matrix}
		Q = \begin{bmatrix}
		R_o & -R_o \\ -R_o & R_o
		\end{bmatrix}.
		\end{align}
	\end{subequations}
	Since $R_o$ is invertible by definition, $Q$ is a positive semi-definite matrix, the minimum is found when $\bm{\chi}^\top Q \bm{\chi}=0$. 
	In order to obtain a non-trivial solution, one has to choose $\bmvarphi_{c}^\top (\x_m) K^\top = \bmvarphi_{o}^\top(\x_m)$, i.e., the added flexibility has to be written as a linear combination of the canonical model.  Hence, adding flexibility to the overparameterized model will always increase \eqref{eq:OverSmallerProof_mini}. This concludes that the calculated uncertainty in the prediction for an overparameterized model with redundant parameters of Category 2 will be larger compared to that of the canonical model.      	
	\begin{flushright}
		\qed
	\end{flushright}	
\end{pf}
Note that choosing $\bmvarphi_{c}^\top(\x_m) K^\top = \bmvarphi_{o}^\top(\x_m)$ would result in that neither \eqref{eq:Larger_model_info} nor $R_o$ is invertible, i.e., the same setting as Theorem~\ref{th:redundentNonlin}, where $T=K^\top$. Hence equality in \eqref{eq:sigVarOverLinpar} can only be obtained when the redundant parameters are of Category 1.  

By formulating a model selection problem, for nested model structures, similar results as Theorem~\ref{th:largerModel} have been shown in \citep{hjalmarsson1995composite} where the calculated uncertainty is higher for the larger model.
That the uncertainty in the prediction is higher for an overparameterized model compared to a canonical model, is also the premise for using model selection algorithms, such as the Bayesian information criteria (\bic), \citep{schwarz1978estimating}.    
This paper provides insights into why the calculated uncertainty has to be strictly larger for the overparameterized model, and how the result generalizes to the case with redundant parameters of Category 1. That is since the minimum of the minimization problem in \eqref{eq:OverSmallerProof_mini} is obtained when the added regressor can be written as a linear combination of the regressor of the canonical model, added flexibility must increase the calculated uncertainty in the prediction.

The results of  Theorem~\ref{th:redundentNonlin} and Theorem~\ref{th:largerModel} can be summarised as for an overparameterized model, the calculated uncertainty from using the delta method will always be larger compared to a canonical model.  However, if the additional parameters do not add any flexibility, the uncertainty in the prediction is the same.
In practice, by observing structure of the chosen model, it is hard to distinguish between the two aforementioned categories of overparameterization. However, after quantifying the uncertainty, if the information matrix $\mathcal{I}^{\bmtheta}$ is rank deficient it would indicate that it is likely that the model might have redundant parameters of Category 1.

\subsection{Nonlinear models}
The main idea of the delta method is a two-step linearization. Firstly, to compute the parameter uncertainty, and secondly to propagate the uncertainty to the output of the model. Hence, the delta method gives a linear approximation of a nonlinear model.  Asymptotically in the number of data points, close to the \mle, a good approximation for many nonlinear models is given by a linear approximation, \citep{nocedal2006numerical,ljung1994modeling,enqvist2005}.     
Consequently, using  \eqref{eq:linearize} to calculate the uncertainty in the prediction for these nonlinear models, the result of Theorem~\ref{th:redundentNonlin} and Theorem~\ref{th:largerModel} should hold asymptotically close to the estimated parameters since the delta method uses a linear approximation of the nonlinear model.

\begin{figure*} [bth!]
	\centering
	\begin{subfigure}[b]{0.32\textwidth}
		\centering
		\includegraphics[width=\textwidth]{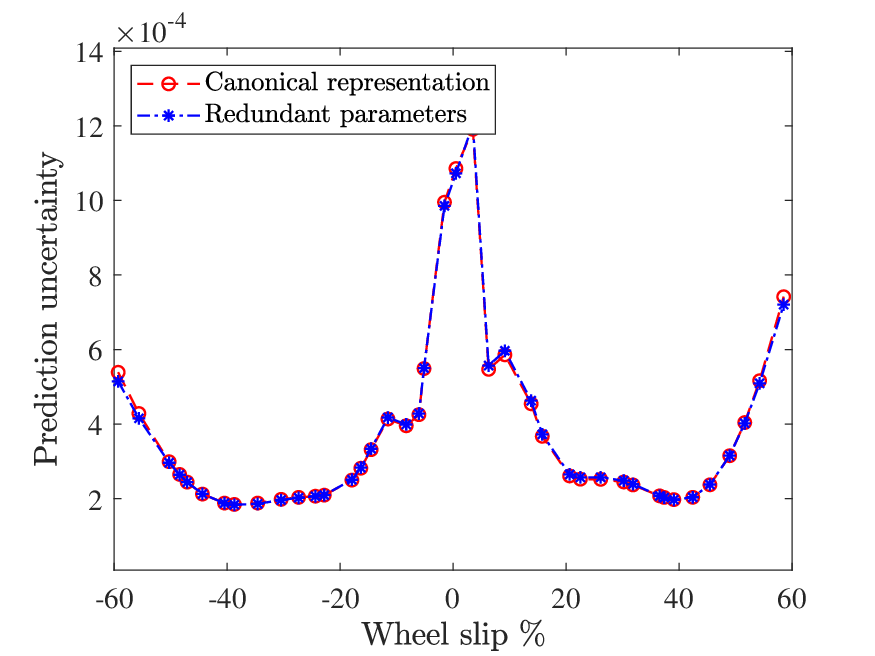}
		\caption{ \footnotesize Nonlinear model with redundant parameters Category 1.}
		\label{fig:redundent}
	\end{subfigure}
	\hfill
	\begin{subfigure}[b]{0.32\textwidth}
		\centering
		\includegraphics[width=\textwidth]{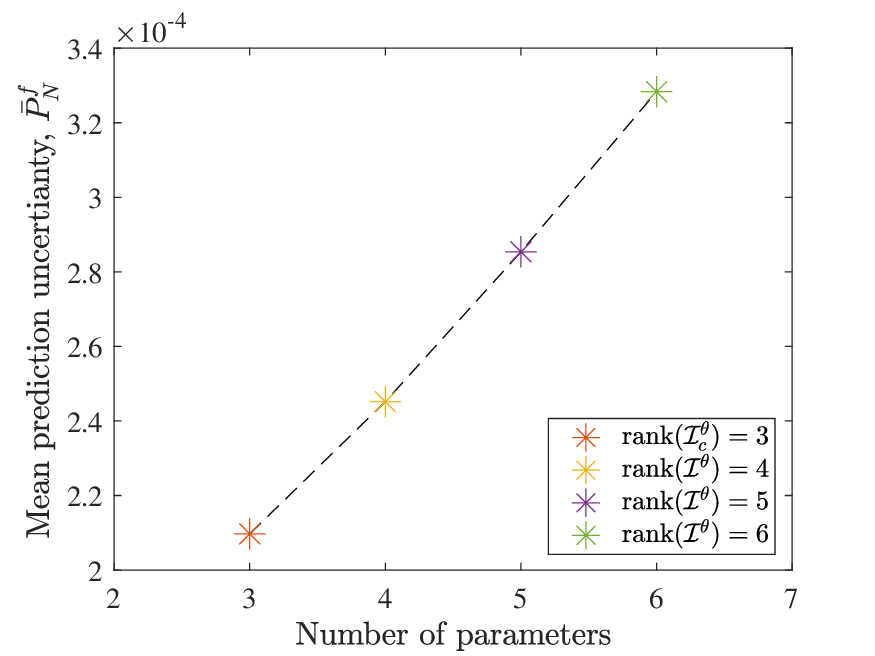}
		\caption{ \footnotesize Linear model with redundant parameters Category 2.}
		\label{fig:moreflex}
	\end{subfigure}
	\hfill
	\begin{subfigure}[b]{0.32\textwidth}
		\centering
		\includegraphics[width=\textwidth]{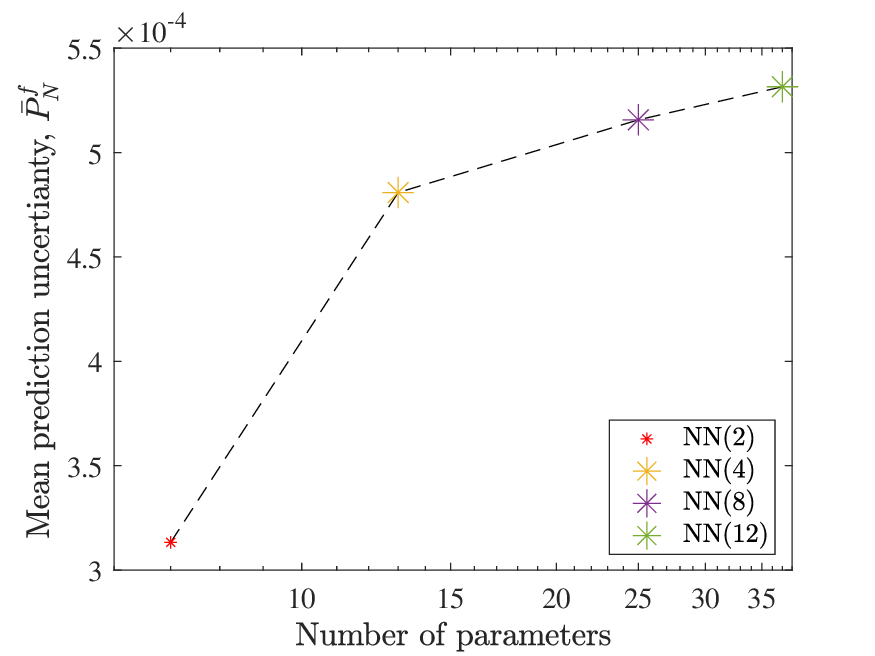}
		\caption{ \footnotesize Nonlinear model with redundant parameters Category 2.}
		\label{fig:moreflex2}
	\end{subfigure}
	\caption{\footnotesize Calculated uncertainty in the prediction for a canonical and an overparameterized model using the delta method. In (a) the model is nonlinear and the overparameterized model has redundant parameters of Category 1. While in (b) the model is linear in its parameters and the overparameterization has redundant parameters of Category 2 which add flexibility to the model. In (c) the model is a two-layer \nn with an increasing number of nodes in the hidden layer, i.e., it is nonlinear and the overparameterization with redundant parameters of Category 2. The simulation data is generated by \eqref{eq:magic}.  } %
	\label{fig:variance_over}
\end{figure*}

\section{Numerical examples} \label{sec:experiment}
\noindent A simulation study will be used to validate the results from Theorem~\ref{th:redundentNonlin} and Theorem~\ref{th:largerModel}. 
The true system under consideration is how the normalized traction force depends on the wheel slip referred to as the \textit{magic formula tire model}, \citep{pacejka1997magic}, 
\begin{align} \label{eq:magic}
f^\ast(x)\! \!= \! \! D \sin \! \big( \! C \! \arctan \big(\! Bx \! -  \! E(Bx \! - \! \arctan(Bx)) \!\big)\!\big).
\end{align}
It is exactly modeled by an \nn with two layers and two nodes in the hidden layer \citep{malmstrom2021}. Hence the true system is included in the model set. 
For the simulation $B =14$, $C=0.1$, $D=0.6$, $E=-0.2$ in \eqref{eq:magic}, $200$ mesurements are generated where $x\sim \mathcal{U} [-0.6,0.6]$, and $e\sim \mathcal{N}(0,0.01)$.

\subsection{Models with redundant parameters of Category 1}
\noindent A  nonlinear canonical model given by 
\begin{align} \label{eq:canon_redudent}
f_c(x, \bmthetac) \! = \! \bmthetac_1 \sigma(\bmthetac_2 x \!+ \! \bmthetac_3)\! + \!\bmthetac_4 \sigma(\bmthetac_5 x \! + \! \bmthetac_6) \!+ \! \bmthetac_7,
\end{align}
and an example of an overparameterized model with redundant parameters of Category 1 is e.g.,  
\begin{align} \label{eq:over_redudent}
f(x, \bmtheta)  = & \bmtheta_1 \sigma(\bmtheta_2 (x+1) \! + \! \bmtheta_3(2-x) \!+ \! 5\bmtheta_4) \nonumber \\
&+\bmtheta_5 \sigma(\bmtheta_6 x + \bmtheta_7) + \bmtheta_8.
\end{align}
The parameters of these models are estimated using \eqref{eq:paramest}, and then the uncertainty in the parameters and prediction is computed using \eqref{eq:linearize}. From Fig.~\ref{fig:redundent} one can conclude that the uncertainty of the overparameterized model with redundant parameters of Category 1 is the same as the uncertainty of the canonical one.

\subsection{Models with redundant parameters of Category 2}
\noindent A canonical model linear in its parameters is given by  
\begin{align} \label{eq:lincanon}
\bmvarphi_{c}(x) \! \! =\! \! [ \sigma(\!-40 x \!\! + \!0.0061), \sigma(\!-6.8 x \! \!+ \! 0.0036), 1 ]^\top \! \!.
\end{align}
And a  model with more flexibility but with redundant parameters of Category 2 is defined by  
\begin{subequations} \label{eq:largermodels}
	\begin{align} 
	\bmvarphi^\top(x) & = [  \bmvarphi_{c}^\top(x), \bmvarphi_{o_j}^\top(x) ], \\
	\bmvarphi_{o_j}^\top(x) & = [
	\sigma(W_{o_j} x + b_{o_j} ), \bmvarphi_{o_{j-1},m}^\top 
	],  
	\end{align}
	where $o_j = 1, ...,3$, and 
	$	W_{o_j} = b_{o_j} = o_j.$
\end{subequations}

A nonlinear canonical model is given by a fully connected two-layer \nn with two nodes in the hidden layer and sigmoid as an activation function, i.e., \eqref{eq:deepnn} with $L=2$. Increasing the number of nodes in the hidden overparameterized model with redundant parameters of Category 2 can be obtained.  

Once again the parameters are estimated using \eqref{eq:paramest} and the uncertainty in the prediction is calculated using \eqref{eq:linearize}, both for the nonlinear model and the model linear in its parameters.
For each model, the mean uncertainty is computed for $N_v$ validation data points according to
\begin{align}
\bar{P}^f_N=\frac{1}{N_v} \sum_{i=1}^{N_v} P^{f}_N (\x_i).
\end{align}

From Fig.~\ref{fig:moreflex} and Fig.~\ref{fig:moreflex2}, Theorem~\ref{th:largerModel} can be verified as the models with redundant parameters of Category 2 have a higher mean prediction uncertainty compared to the canonical model.
  
\section{Conclusion}
\noindent  
When calculating the uncertainty in the prediction using the delta method, for an overparameterized model, the prediction uncertainty will always be larger or equal compared to one of the models with minimum flexibility that still can describe the true system, i.e., a canonical model.
Hence, the conclusion is that even though the model is overparameterized, the uncertainty in the prediction calculated by the delta method is not underestimated, i.e., the uncertainty in the prediction is not too low.
The delta method relies on linearizations, hence asymptotically in the number of data points, the results from Theorem~\ref{th:redundentNonlin} and Theorem~\ref{th:largerModel} apply locally around the estimated parameters for some nonlinear models. Section~\ref{sec:experiment} provides an example of two such nonlinear models.

A future research direction could be to investigate if the result also holds for larger black-box models such as \nn{s} used for image classification.

\addtolength{\textheight}{-12.1cm}

\bibliography{IEEEabrv,mybibref5G}                                                                                                      

\end{document}